\documentclass{aipproc}

\layoutstyle{6x9}

\SetInternalRegister\hbadness{8000} 

\newcommand\doingARLO[2][]{%
  \ifx\mmref\undefined #1\else #2\fi
}

\begin{document}

\title{Division Algebras and Extended SuperKdVs.}


\author{F. Toppan}{
  address={CBPF, Rua Dr. Xavier Sigaud 150, cep 22290-180 Rio de Janeiro (RJ),
  Brazil},
  email={toppan@cbpf.br},
}

\copyrightyear{2001}

\begin{abstract}
The division algebras ${\bf R}$, ${\bf C}$, {\bf H}$,
{\bf O}$ are used to construct and analyze the $N=1,2,4,8$ supersymmetric
extensions of the KdV hamiltonian equation. In particular
a global $N=8$ super-KdV system is introduced and shown to admit a
Poisson bracket structure given by the ``Non-Associative $N=8$
Superconformal Algebra".

\end{abstract}

\date{\today}

\maketitle

\section{Introduction}

In the last several years integrable hierarchies of non-linear
differential equations in $1+1$ dimensions have been intensely
explored, mainly in connection with the discretization of the
two-dimensional gravity (see \cite{toppDGZ:1995}).\par
Supersymmetric extensions of such equations have also been largely
investigated \cite{toppMR:1985}-\cite{toppT:1995} using a variety
of different methods. Unlike the bosonic theory, many questions
have not yet been answered in the supersymmetric case. In this
talk I report a recent result, obtained in collaboration with H.L.
Carrion and M. Rojas \cite{toppCRT2:2001}, concerning the
formulation of the $N$-extended supersymmetric versions of the
bosonic integrable equations (for $N=1,2,4,8$) in terms of the
division algebras ${\bf R}$, ${\bf C}$, {\bf H}$, {\bf O}$
respectively.\par To be precise we focused our investigation on
the supersymmetric extensions of KdV. At first the standard
results of Mathieu \cite{toppM:1988} concerning $N=2$ KdV are
reviewed in the language of division algebras, while a full
analysis of the global $N=4$ extensions of KdV is performed. The
Delduc and Ivanov \cite{toppDI:1993} result is recovered as a
special case. Later it is proven that a unique $N=8$  hamiltonian
extension of KdV can be found. It admits as a Poisson brackets
structure the so-called ``Non-Associative $N=8$ Superconformal
Algebra" introduced for the first time in \cite{toppESSTP:1988}.
It is quite a special superconformal algebra, since it does not
satisfy the Jacobi property (this is why it was named
``non-associative"). In order to have an $N=8$ KdV, the
``non-associativity" of the underlining superconformal algebra of
Poisson brackets is mandatory. This is due to the fact that no
central extension, which is responsible for the inhomogeneous
character of KdV, is allowed by jacobian superconformal algebras.
\par
It is worth noticing that in the following the problem of the integrability of the
hamiltonian super-KdVs system is not addressed, only the issue of their global
supersymmetric invariances is of concern here.


\section{Division Algebras and Extended Superconformal Algebras}

The $N$-extended superconformal algebras, for $N=1,2,4,8$, can be
recovered from division algebras. Let us here present the largest
of such conformal algebras, the ``$N=8$ Non-associative SCA" of
reference \cite{toppESSTP:1988}. The $N=1,2,4$ SCA's are recovered
as subalgebras. The ``$N=8$ Non-associative SCA" can be defined
via octonionic structure constants.
 A generic octonion $x$ is
expressed as $x=x_a\tau_a$ (throughout the text the convention
over repeated indices is understood), where $x_a$ are real numbers
while $\tau_a$ denote the basic octonions, with
$a=0,1,2,...,7$.\par $\tau_0\equiv {\bf 1}$ is the identity, while
$\tau_\alpha$, for $\alpha =1,2,...,7$, denote the imaginary
octonions. In the following a Greek index is employed for
imaginary octonions, a Latin index for the whole set of octonions
(identity included).\par The octonionic multiplication can be
introduced through
\begin{eqnarray}
\tau_\alpha \cdot \tau_\beta &=& -\delta_{\alpha\beta} \tau_0 +
C_{\alpha\beta\gamma} \tau_\gamma , \label{topp-eq1}
\end{eqnarray}
with $C_{\alpha\beta\gamma}$ a set of totally antisymmetric
structure constants which, without loss of generality, can be
taken to be
\begin{eqnarray}
&C_{123}=C_{147}=C_{165}=C_{246}=C_{257}= C_{354}=C_{367}=1.&
\label{topp-eq2}
\end{eqnarray}
and vanishing otherwise.\par When $\alpha,\beta,\gamma$ are
restricted to, let's say, the values $1,2,3$ we recover the
quaternionic subalgebra, which is associative. The $N=8$ extension
of the Virasoro algebra is constructed in terms of the above
structure constants. Besides the spin-$2$ Virasoro field, it
contains eight fermionic spin-$\frac{3}{2}$ fields $Q$, $Q_\alpha$
and $7$ spin-$1$ bosonic currents $J_\alpha$. It is explicitly
given by the following Poisson brackets
\begin{eqnarray}
\{ T(x), T(y)\} &=& -\frac{1}{2} {\partial_y}^3 \delta(x-y) +
2T(y)\partial_y\delta(x-y) +T'(y) \delta(x-y),\nonumber\\ \{T(x),
Q(y)\} &=& \frac{3}{2}Q(y)\partial_y\delta (x-y) + Q'(y) \delta
(x-y)+ (X1),\nonumber\\ \{T(x), Q_\alpha(y)\} &=&
\frac{3}{2}Q_{\alpha}(y)\partial_y\delta (x-y) + {Q_\alpha}'(y)
\delta (x-y),\nonumber\\
 \{T(x), J_\alpha(y)\} &=&
J_{\alpha}(y)\partial_y\delta (x-y) + {J_\alpha}'(y) \delta
(x-y),\nonumber\\
 \{Q(x), Q(y)\} &=&-\frac{1}{2}{\partial_y}^2\delta(x-y)+
+\frac{1}{2} {T}(y) \delta (x-y),\nonumber\\
 \{Q(x), Q_\alpha(y)\} &=&
-J_{\alpha}(y)\partial_y\delta (x-y) -\frac{1}{2} {J_\alpha}'(y)
\delta (x-y),\nonumber\\
 \{Q(x), J_\alpha(y)\} &=&
-\frac{1}{2}Q_{\alpha}(y)\delta (x-y),\nonumber\\
 \{Q_\alpha(x), Q_\beta(y)\}
 &=&-\frac{1}{2}\delta_{\alpha\beta}{\partial_y}^2\delta(x-y) +
 C_{\alpha\beta\gamma}
 J_\gamma(y)\partial_y\delta(x-y)+\nonumber\\&& +
\frac{1}{2}(\delta_{\alpha\beta}T(y)+C_{\alpha\beta\gamma}
{J_\gamma}'(y))\delta(x-y),\nonumber\\
 \{Q_\alpha(x), J_\beta(y)\} &=&
\frac{1}{2}(\delta_{\alpha\beta} Q(y)-C_{\alpha\beta\gamma}
Q_\gamma (y)) \delta(x-y),\nonumber\\
 \{J_\alpha(x), J_\beta(y)\} &=&
\frac{1}{2}\delta_{\alpha\beta}\partial_y\delta (x-y) -
C_{\alpha\beta\gamma} J_\gamma(y)\delta (x-y). \label{topp-eq3}
\end{eqnarray}
This superconformal algebra can be recovered via Sugawara
construction of the affine octonionic algebra (see
\cite{toppCRT1:2001} for details). The failure in closing the
Jacobi identity is in consequence of the non-associativity of the
multiplication between octonions.

\section{Extended SuperKdVs}

A natural question to be raised is whether the (\ref{topp-eq3})
SCA (and its superconformal subalgebras) can be regarded as a
Poisson brackets structure for a supersymmetric hamiltonian
extension of KdV. This amounts to determine the most general
globally supersymmetric hamiltonian of a given dimension $4$ (i.e.
the second hamiltonian). In the case of $N=2$, this result is
known since the works of Mathieu \cite{toppM:1988}. In
\cite{toppCRT2:2001} the extension to the $N=4$ case has been
completely worked out. The moduli space of inequivalent $N=4$ KdV
equations has been classified. The special integrable point of
Delduc-Ivanov \cite{toppDI:1993} has been recovered in this
framework. The complete solution to the $N=8$ case was given as
well.\par To get these results an extensive use of computer
algebra (with Mathematica and the Thielemans' package for
classical OPEs computations) was required. The final results are
presented here.\par The most general $N=4$-supersymmetric
hamiltonian for the $N=4$ KdV depends on $5$ parameters (plus an
overall normalization constant). However, if the hamiltonian is
further assumed to be invariant under the involutions of the $N=4$
minimal SCA, three of the parameters have to be set equal to $0$.
I recall here that the involutions of the $N=4$ minimal SCA are
induced by the involutions of the quaternionic algebra. Three such
involutions exist (any two of them can be assumed as generators),
the $\alpha$-th one (for $\alpha=1,2,3$) is given by leaving
$\tau_\alpha$ (together with the identity) invariant and flipping
the sign of the two remaining $\tau$'s.\par What is left is the
most general hamiltonian, invariant under $N=4$ and the
involutions of the algebra. It is given by
\begin{eqnarray}
H_2 &=& T^2 +Q_a'Q_a -J_\alpha''J_\alpha +x_\alpha T{J_\alpha}^2
+\nonumber\\ && 2x_\alpha Q_0Q_\alpha J_\alpha
-C_{\alpha\beta\gamma}x_\alpha J_\alpha Q_\beta Q_\gamma
+\nonumber\\ && 2 x_1J_1J_2'J_3-2x_2J_1'J_2J_3. \label{topp-eq4}
\end{eqnarray}
here $a=0,1,2,3$ ($Q_0\equiv Q$) and $\alpha,\beta,\gamma \equiv
1,2,3$.
\par
The $N=4$ global supersymmetry requires the three parameters $x_a$
to satisfy the condition
\begin{eqnarray}
x_1+x_2+x_3&=&0, \label{topp-eq5}
\end{eqnarray}
so that only two of them are independent. Since any two of them,
at will, can be plotted in a real $x-y$ plane, it can be proven
that the fundamental domain of the moduli space of inequivalent
$N=4$ KdV equations can be chosen to be the region of the plane
comprised between the real axis $y=0$ and the $y=x$ line
(boundaries included). There are five other regions of the plane
(all such regions are related by an $S_3$-group transformation)
which could be equally well chosen as fundamental domain.\par In
the region of our choice the $y=x$ line corresponds to an extra
global $U(1)$-invariance, while the origin, for $x_1=x_2=x_3=0$,
is the most symmetric point (it corresponds to a global $SU(2)$
invariance associated to the generators $\int dx J_\alpha (x)$).
\par
The involutions associated to each given imaginary quaternion
allows to consistently reduce the $N=4$ KdV equation to an $N=2$
KdV, by setting simultaneously equal to $0$ all the fields
associated with the $\tau$'s which flip the sign, e.g. the fields
$J_2=J_3=Q_2=Q_3=0$ for the first involution (and similarly for
the other couples of values $1,3$ and $1,2$). After such a
reduction we recover the $N=2$ KdV equation depending on the free
parameter $x_1$ (or, respectively, $x_2$ and $x_3$).
\par The integrability is known for $N=2$ KdV to be ensured for three
specific values $a=-2,1,4$, discovered by Mathieu
\cite{toppM:1988}, of the free parameter $a$. We are therefore in
the position to determine for which points of the fundamental
domain the $N=4$ KdV is mapped, after any reduction, to one of the
three Mathieu's integrable $N=2$ KdVs. It turns out that in the
fundamental domain only two such points exist. Both of them lie on
the $y=x$ line. One of them produces, after inequivalent $N=2$
reductions, the $a=-2$ and the $a=4$ $N=2$ KdV hierarchies. The
second point, which produces the $a=1$ and the $a=-2$ $N=2$ KdV
equations, however, does not admit at the next order an $N=4$
hamiltonian which is in involution w.r.t. $H_2$. This has been
explicitly proven in \cite{toppCRT2:2001}. The treatment of
\cite{toppCRT2:2001} is more complete than the one in
\cite{toppDI:1993} since it is based on an exhaustive
component-fields analysis, rather than on an extended superfield
formalism.\par The most general equations of motion of the $N=4$
KdV are directly obtained from the hamiltonian (\ref{topp-eq4})
together with the (\ref{topp-eq3}) Poisson brackets.

\section{The $N=8$ SuperKdV}

A similar analysis can be extended to the $N=8$ case based on the
full $N=8$ non-associative SCA. At first the most general
hamiltonian with the right dimension has been written down. Later,
some constraints on it have been imposed. The first set of
constraints requires the invariance under all the $7$ involutions
of the algebra. In the case of octonions the total number of
involutions is $7$ (with $3$ generators) each one being associated
to one of the seven combinations appearing in (\ref{topp-eq2}). In
the case, e.g., of the $123$ combination the corresponding
$\tau_\alpha$'s are left invariant, while the remaining four
$\tau$'s, living in the complement, have the sign flipped.
\par The second set of constraints requires the invariance under
the whole set of $N=8$ global supersymmetries. Under this
condition there exists only one hamiltonian, up to the
normalization factor, which is $N=8$ invariant. It does not
contain any free parameter and is quadratic in the fields. It is
explicitly given by
\begin{eqnarray}
H_2 &=& T^2 + Q_a'Q_a - J_\alpha''J_\alpha \label{topp-eq6}
\end{eqnarray}
(here $a=0,1,...,7$ and $\alpha=1,2,...,7$).\par The hamiltonian
corresponds to the origin of coordinates (confront the previous
case) which is also, just like the $N=4$ case, the point of
maximal symmetry. This means that the hamiltonian is invariant
under the whole set of seven global charges $\int dx J_\alpha(x)$,
obtained by integrating the currents $J_\alpha$'s.
\par Despite its apparent simplicity, it gives
an $N=8$ extension of KdV which does not reduce (for any $N=2$
reduction) to the three Mathieu's values for integrability.
Nevertheless it is a highly non-trivial fact that an $N=8$
extension of the KdV equation indeed exists and that it is unique.
Explicitly, the associated equations of motion are given by
\begin{eqnarray}
{\dot T} &=& T''' + 12 T' T + 6 Q_a'' Q_a -4
J_\alpha''J_\alpha,\nonumber\\ {\dot Q} &=& Q''' +6 T' Q +6 T Q' +
4 Q_\alpha''J_\alpha - 2 Q_\alpha J_\alpha'',\nonumber\\ {\dot
Q}_\alpha &=& {Q_\alpha}''' + 2 Q J_\alpha'' + 6 TQ_\alpha' + 6 T'
Q_\alpha - 2 Q' J_\alpha' -4 Q''J_\alpha + \nonumber\\ && +
C_{\alpha\beta\gamma} ( Q_\beta J_\gamma'' -Q_\beta' J_\gamma' -2
Q_\beta''J_\gamma ),\nonumber\\ {\dot J}_\alpha &=& {J_\alpha}'''
+ 4 T'J_\alpha + 4 T J_\alpha ' + 2 Q Q_\alpha '
+C_{\alpha\beta\gamma}( 2 J_\alpha J_\beta J_\gamma'' +Q_\beta
Q_\gamma' ). \label{topp-eq7}
\end{eqnarray}
Besides the $N=8$ KdV, the fields entering (\ref{topp-eq3}) admit
globally $N=4$ supersymmetric invariant hamiltonians, which depend
on free parameters. Two such classes of hamiltonians are
individuated. The first class consists of the hamiltonians
invariant under supersymmetries related with the quaternionic
subalgebra which, without loss of generality, can be assumed to be
given by $\int dx Q_i(x)$, for $i=0,1,2,3$. The second class of
invariances is associated to the $N=4$ supersymmetries associated
to the remaining generators, i.e. those living in the complement
(in the following, without loss of generality, these
supersymmetries are labeled by $i=1,2,4,5$). For completeness we
report here the results. The first class of $N=4$-invariant
hamiltonians is given by
\begin{eqnarray}
H_2 &=& T^2 + Q'Q +Q_p'Q_p  + x Q_r'Q_r -J_p''J_p -x J_r''J_r +2
x_p Q Q_pJ_p +x_p T J_pJ_p-\nonumber\\ &&- 2 x_3 Q_1Q_2J_3 +2 x_2
Q_1Q_3J_2 -2 x_1 Q_2Q_3 J_1 +2 x_1 J_1J_2'J_3-2 x_2 J_1'J_2J_3.
\label{topp-eq8}
\end{eqnarray}
with $x_1$, $x_2$, $x$ free parameters, while $x_3=-x_1-x_2$.
\par In the above expression $p=1,2,3$ and $r=4,5,6,7$.\par The second
class of $N=4$ invariant hamiltonians is explicitly given by
\begin{eqnarray}
H_2 &=& T^2 + x (Q'Q  +Q_3'Q_3+Q_6'Q_6 +Q_7'Q_7) +
Q_1'Q_1+Q_2'Q_2+Q_4'Q_4+Q_5'Q_5 - \nonumber\\ && -x( J_1''J_1
+J_2''J_2+ J_4''J_4+J_5''J_5) -J_3''J_3-J_6''J_6-J_7''J_7
+\nonumber\\ &&+ y (TJ_7J_7 - TJ_3J_3) + 2y J_3J_6'J_7 +
2y(Q_1Q_2J_3-Q_1Q_4J_7-Q_2Q_5J_7-Q_4Q_5J_3) \nonumber\\ &&
\label{topp-eq9}
\end{eqnarray}
where in this case two free parameters, $x$ and $y$, appear.

\section{CONCLUSIONS}

In this paper I have presented some new results concerning an
explicit connection of division algebras and the extended
supersymmetrizations of the KdV equation. \par In particular it
has been proven, following \cite{toppCRT2:2001}, the existence of
a unique $N=8$ KdV equation of hamiltonian type based on the $N=8$
non-associative SCA as a generalized classical Poisson brackets
structure.
\par Division algebras are a natural ingredient when dealing with
extended supersymmetries. It is therefore likely that
supersymmetric extensions of other classes of bosonic integrable
equations could be studied with the tools furnished by division
algebras. Besides KdV, the next simplest equations to be
investigated are the mKdV and the NLS. In view of the results of
\cite{toppCRT1:2001}, where the $N=8$ Non-associative SCA is
recovered from a singular limit of a Sugawara construction based
on the superaffine octonionic algebra of superMalcev type, it is
almost for granted that such extensions indeed exist.\par Whether
such constructions could be applied to other classes of integrable
equations is still an open problem. In any case division algebras
look as a promising and elegant tool to unveil some of the
mysterious features still surrounding the supersymmetrization of
the bosonic hierarchies.

\begin{theacknowledgments}
I am grateful to the organizers of the Karpacz Winter School for
the invitation. It is a pleasure for me to thank my collaborators
H.L. Carrion and M. Rojas.
\end{theacknowledgments}


\begin{thebibliography}{8}
\expandafter\ifx\csname natexlab\endcsname\relax\def\natexlab#1{#1}\fi
\providecommand{\enquote}[1]{``#1''}
\expandafter\ifx\csname url\endcsname\relax
  \def\url#1{\texttt{#1}}\fi
\expandafter\ifx\csname urlprefix\endcsname\relax\def\urlprefix{URL }\fi

\bibitem{toppDGZ:1995}
Di Francesco, P., Ginsparg P., and Zinn-Justin, J., \emph{Phys.
Rep.}, \textbf{254}, 1995, pp. 1.

\bibitem{toppMR:1985}
Manin, Y.I., and Radul, A.O., \emph{Comm. Math. Phys.}
\textbf{98}, 1985, pp. 65.

\bibitem{toppM:1988}
Mathieu, P., \emph{Phys. Lett.}, \textbf{B 203}, 1988, pp. 65;
Laberge, C.A., and Mathieu, P., \emph{Phys. Lett.} \textbf{B 215},
1988, pp. 718; Labelle, P., and Mathieu, P., \emph{J. Math. Phys.}
\textbf{89}, 1991, pp. 923.

\bibitem{toppIK:1991}
Inami, T., and Kanno, H., \emph{Comm. Math. Phys.}, \textbf{136},
1991, pp. 519; \emph{Int. J. Mod. Phys.} \textbf{A 7}, Suppl. 1A
1992, pp. 419.

\bibitem{toppP:1994}
Popowicz, Z., \emph{Phys. Lett.}, \textbf{A 194}, 19994, pp. 375;
\emph{J. Phys.} \textbf{A 29}, 1996, pp. 1281; \emph{ibid.}
\textbf{39}, 1997, pp. 7935; \emph{Phys. Lett.} \textbf{B 459},
1999, pp. 150.

\bibitem{toppBD:1995} Brunelli, J.C., and Das, A., \emph{J. Math.
Phys.} \textbf{36}, 1995, pp. 268.

\bibitem{toppT:1995} Toppan, F., \emph{Int. J. Mod.
Phys.}\textbf{A 10}, 1995, pp. 895.


\bibitem{toppCRT2:2001} Carrion, H.L., Rojas, M., and Toppan, F.,
Division Algebras and the N=1,2,4,8 Extensions of KdV, Preprint
CBPF-NF-012/01, 2001.


\bibitem{toppDI:1993} Delduc, F., and Ivanov, E., {\emph Phys.
Lett.} \textbf{B 309}, 1993, pp. 312; Delduc, F., Ivanov, E., and
Krivonos, S., \emph{J. Math. Phys.} \textbf{37}, 1996, pp. 1356.

\bibitem{toppESSTP:1988} Englert, F., Sevrin, A., Troost, W., van
Proeyen, A., and Spindel, P., \emph{J. Math. Phys.} \textbf{29},
1988, pp. 281.

\bibitem{toppCRT1:2001} Carrion, H.L., Rojas, M., and Toppan, F.,
An N=8 Superaffine Malcev Algebra and Its N=8 Sugawara, Preprint
CBPF-NF-011/01, 2001.


\end{thebibliography}
\end{document}